\documentclass[twocolumn,aps,prl,superscriptaddress]{revtex4-2}
\usepackage{amsmath,amssymb,amsfonts}
\usepackage{bm}
\usepackage{graphicx}
\usepackage{mathrsfs}
\usepackage{multirow}
\usepackage[normalem]{ulem}
\usepackage{graphicx}
\usepackage{booktabs, ctable}
\usepackage{xcolor, color, framed}
\usepackage[colorlinks, linkcolor=red,anchorcolor=green,citecolor=blue]{hyperref}

\begin{document}

\title{D-meson production via sequential hadronization in high-energy nuclear collisions}

\date{\today  \hspace{1ex}}

\author{Zi-Xuan Xu}
\email{zixuan@mails.ccnu.edu.cn}
\affiliation{Key Laboratory of Quark \& Lepton Physics (MOE) and Institute of Particle Physics, Central China Normal University, Wuhan 430079, China}

\author{Wei Dai}
\email{corresponding author: weidai@cug.edu.cn}
\affiliation{School of Mathematics and Physics, China University of Geosciences, Wuhan 430074, China }

\author{Ben-Wei Zhang}
\email{bwzhang@mail.ccnu.edu.cn}
\affiliation{Key Laboratory of Quark \& Lepton Physics (MOE) and Institute of Particle Physics, Central China Normal University, Wuhan 430079, China}

\author{Jiaxing Zhao}
\email{jzhao@itp.uni-frankfurt.de}
\affiliation{Helmholtz Research Academy Hesse for FAIR (HFHF), GSI Helmholtz Center for Heavy Ion Physics, Campus Frankfurt, 60438 Frankfurt, Germany}
\affiliation{Institut f\"ur Theoretische Physik, Johann Wolfgang Goethe-Universität, Max-von-Laue-Straße 1, D-60438 Frankfurt am Main, Germany}

\author{Pengfei Zhuang}
\email{zhuangpf@mail.tsinghua.edu.cn}
\affiliation{Department of Physics, Yantai University, Yantai 264005, China}
\affiliation{Department of Physics, Tsinghua University, Beijing 100084, China}
\affiliation{South Center for Nuclear-Science Theory, Institute of Modern Physics, Chinese Academy of Sciences, Huizhou 516000, China}

\begin{abstract}
Heavy flavor production serves as an ideal probe of the hadronization mechanism of the quark-gluon plasma created in relativistic heavy ion collisions. We study charm-quark hadronization using Langevin transport in the medium together with a sequential coalescence model. Since $D_s$ forms earlier than $D^0$, as obtained from the Dirac equation with an in-medium potential extracted from lattice QCD, the $D_s$ elliptic flow $v_2$ is smaller than the $D^0$ $v_2$ in the intermediate-$p_T$ region, in good agreement with the recent ALICE data. Incorporating sequential coalescence, charm-quark number conservation, and strangeness enhancement predicts a peak in the yield ratio $D_s/D^0$ at low $p_T$, which can be tested in future heavy-ion collisions. 
\end{abstract}

\maketitle

\emph{Introduction.--}
It is widely accepted that relativistic heavy-ion collisions can create a deconfined state of Quantum Chromodynamics (QCD) matter, known as the quark-gluon plasma (QGP). As the QGP expands and cools down, it hadronizes via the coalescence mechanism~\cite{Fries:2003kq,Fries:2008hs,Greco:2003mm}. Almost all hadronization models assume that all quarks, including both light and heavy flavors, hadronize simultaneously on a hypersurface determined by the hydrodynamic evolution of the QGP~\cite{Minissale:2020bif,Plumari:2017ntm,Cao:2015hia,Cao:2019iqs,Gossiaux:2009mk,Nahrgang:2013xaa,Zhao:2024ecc,Zhao:2024oma,Song:2015sfa,Song:2015ykw,Ravagli:2007xx,He:2019tik,He:2012df,Andronic:2021erx,Beraudo:2014boa,Beraudo:2017gxw}. However, because quarkonia, which are composed of a heavy-quark pair, can survive in the QGP due to their large binding energies and thus serve as probes of QGP formation~\cite{STAR:2025imj,CMS:2012gvv,Du:2015wha}, heavy quarks should in principle hadronize before light quarks, or at a temperature above the phase transition temperature $T_c$~\cite{Zhao:2018jlw,Zhao:2020rrk}. 
Additionally, a flavor hierarchy in the deconfinement phase transition of light and strange quarks is proposed based on lattice simulations~\cite{Bellwied:2013cta}.
For heavy flavor, the hadronization time (or temperature) can be studied using relativistic potential models with lattice-QCD–motivated heavy-quark potentials, such as two- and three-body Dirac equations~\cite{Crater:1983ew,Crater:1987hm,Shi:2013rga,Shi:2019tji}. This framework, rooted in Dirac’s constraint dynamics, is consistent with the Bethe–Salpeter equation in the two-body limit~\cite{Sazdjian:1986aw,Sazdjian:1988be}.

Qualitatively, sequential hadronization leads to two distinct phenomena. 1) Anomalous yield ratio. At RHIC and LHC energies, heavy quarks are produced through initial hard processes, and the thermal production in the QGP can be neglected. This means heavy quark number conservation during the evolution of the QGP, and there will be more heavy quarks participating in the hadronization at earlier times. As a consequence, the hadrons formed earlier are enhanced, and those formed later are suppressed. This yield ratio induced by the quark number conservation in the sequential hadronization is very different from the simultaneous hadronization, where the number conservation does not change the relative yields. 2) Anomalous flow sequence of charm mesons. The hadrons formed earlier carry less collective flow absorbed from the heavy quark interaction with the QGP, in comparison with the hadrons formed later. For instance, the elliptic flow $v_2(J/\psi) < v_2(D)$ is measured experimentally and explained theoretically at RHIC and LHC energies~\cite{ALICE:2020pvw}. The recent high-precision ALICE measurement of $v_2(D_s^+) < v_2(D^0)$~\cite{ALICE:2026zcz} in the intermediate transverse-momentum region, shown in Fig.~\ref{fig4}, contradicts almost all transport-model predictions based on simultaneous hadronization~\cite{Zhao:2023nrz}. However, once the earlier formation of $D_s$ than $D^0$ is taken into account, this observation can be naturally understood within the framework of sequential hadronization.

The purpose of this Letter is to quantitatively calculate the $v_2$ and yield splitting of $D$ mesons and identify evidence for sequential hadronization. The heavy quark transport in QGP is described by a well-established Langevin equation, the QGP evolution is characterized by an often used hydrodynamics, and the formation time and static properties of heavy flavor hadrons are controlled by the 2- and 3-body Dirac equation.     

\emph{Heavy flavor transport.--}
Initially produced heavy quarks in heavy-ion collisions can be treated within perturbative QCD. We take here the FONLL approach~\cite{Cacciari:2012ny} for the momentum distribution and the Glauber Monte Carlo approach~\cite{Miller:2007ri} for the spatial distribution. The in-medium evolution is simulated with a recently improved Langevin transport model that successfully describes the main features of heavy-flavor jet modification~\cite{Cao:2013ita,Wang:2019xey,Dai:2018mhw,Wang:2020ukj,Wang:2020qwe}. The model incorporates both elastic scattering processes and inelastic gluon radiation in the hot and dense medium, accounting for heavy-quark energy loss. The heavy quark position and momentum evolution are controlled by the Langevin equations,
\begin{eqnarray}
	d{\bm x}/dt & = & {\bm p}/E,\nonumber \\ 
	d{\bm p}/dt & = & -\Gamma {\bm p}+\vec \xi+{\bm p}_g, 
\label{langevin}
\end{eqnarray}
where $E$ is the heavy quark energy, $\Gamma{\bm p}$ is the drag force, $\vec \xi$ represent the noise terms, which is assumed to be a momentum-independent white noise and satisfies the relation $\langle \xi^i(t) \xi^j(t') \rangle=\kappa \delta^{ij}\delta(t-t')$. $\kappa$ is the diffusion coefficient which satisfies the fluctuation-dissipation relation $\kappa=2\Gamma ET$~\cite{Kubo:1966fyg} and is related to the spatial charm quark diffusion coefficient ${\cal D}_s=2T^2/\kappa$. We adopt a constant spatial diffusion coefficient of $2\pi T{\cal D}_s=2.5$, which is given by a recent 2+1 flavor lattice calculation~\cite{Altenkort:2023oms}. The Higher-Twist approach~\cite{Guo:2000nz,Zhang:2003wk,Zhang:2004qm,Majumder:2009ge} is employed to simulate the medium-induced gluon radiation to give the gluon momentum ${\bm p}_g$. The gluon radiation spectrum is
\begin{equation}
    \dfrac{dN}{dxdk^2_{\perp}dt}=\dfrac{2\alpha_sC_sP(x)\hat{q}}{\pi k^4_{\perp}}\sin^2(\dfrac{t-t_i}{2\tau_f})(\dfrac{k^2_{\perp}}{k^2_{\perp}+x^2M^2})^4, 
\label{gluon}
\end{equation}
where $x$ and $k_{\perp}$ are the gluon energy fraction and transverse momentum taken from the mother parton. $C_s$ is the quadratic Casimir in color representation. $P(x)$ is the quark splitting function in vacuum~\cite{Deng:2009ncl}. $\tau_f=2Ex(1-x)/(k^2_{\perp}+x^2M^2)$ denotes the formation time of the gluon, $\hat{q}$ is the jet transport parameter in QGP which can be phenomenologically assumed to be 
$\hat{q}(\tau,r)=\hat{q}_0(T/T_0)^3 (p_{\mu}u^{\mu})/p^0$~\cite{Xie:2019oxg}. We take here $\hat{q}_0 = 1.5{~\rm GeV^2/fm}$ at LHC energy, referring to the $\chi^2$ analysis on single-hadron and dihadron nuclear modification~\cite{Xie:2019oxg}. 

The QGP produced in relativistic heavy-ion collisions is a strongly coupled medium whose evolution can be described by viscous hydrodynamics, providing the temperature and flow fields relevant for heavy-quark evolution. We employ here the (3+1)D CLVisc framework~\cite{Pang:2018zzo,Wu:2021fjf} for the QGP evolution, which has successfully described light-hadron production and collective flow over a wide range of collision energies~\cite{Pang:2018zzo,Wu:2021fjf}.

\emph{Sequential hadronization.--} 
A hybrid approach combining coalescence and fragmentation is the state-of-the-art framework for describing hadronization in the QGP. In coalescence, a heavy quark combines with nearby light quarks in phase space to form a low-momentum hadron. At high transverse momentum $p_T$, hadronization is instead dominated by fragmentation, which is assumed to be the same as in vacuum and is described by fragmentation functions. For a charm quark with momentum $p_c$, the probability to coalesce into a hadron is obtained by integrating the charm and light-quark distributions with the corresponding Wigner function,
\begin{eqnarray}
\label{coalescence}
{dN\over d^3{\bf P}}&=&g \int \prod_{i=1}^n{d^3x_id^3p_i\over (2\pi)^3E_i} f_i({\bf x}_i,{\bf p}_i)\\
&\times& W({\bf x}_1,\cdots,{\bf x}_n, {\bf p}_1,\cdots, {\bf p}_n)\, \delta^{(3)}\left({\bf P}-\sum_{i=1}^k{\bf p}_i\right),\nonumber
\end{eqnarray}   
where $g$ is the statistical factor for the spin and color degrees of freedom, $n$ is $2$ for mesons and $3$ for baryons, and $f_i(x_i,p_i)$ is the quark phase-space distribution. The Wigner function $W$ is the quasi-probability for $n$ quarks to form a hadron. In principle, it can be calculated from the hadron wavefunction. For the ground and low excitation states, their Wigner functions are well approximated by harmonic oscillators with suitably chosen widths~\cite{Zhao:2025cnp}. For the charmed-meson states from $1S$ to $2P$ considered here, $W$ can be expressed as~\cite{Zhao:2025cnp}
\begin{eqnarray}
W_{1S}({\bm r},{\bm p})&=&8e^{-\xi},\nonumber\\
W_{1P}({\bm r},{\bm p})&=&\dfrac{8}{3}e^{-\xi}(2\xi - 3),\nonumber\\
W_{1D}({\bm r},{\bm p})&=&\dfrac{8}{15}e^{-\xi}(15+4\xi^2-20\xi+8\eta),\nonumber\\
W_{2S}({\bm r},{\bm p})&=&\dfrac{8}{3}e^{-\xi}(3+2\xi^2-4\xi-8\eta),\nonumber\\         
W_{2P}({\bm r},{\bm p})&=&\dfrac{8}{15}e^{-\xi}(-15+4\xi^3-22\xi^2+30\xi \nonumber\\ 
&&-8(2\xi-7)\eta)
\label{wigner}
\end{eqnarray}
with $\xi = r^2/\sigma^2+p^2\sigma^2$ and $\eta= p^2r^2-({\bf p}\cdot {\bf r})^2$, where ${\bm r}$ and ${\bm p}$ denote the relative coordinate and momentum of the two quarks. For charmed baryons, we first combine two quarks into a diquark and then couple it with the third quark. Consequently, the corresponding Wigner function can be expressed as the product of two Gaussian functions. The width parameter $\sigma$ in the Wigner function can be taken as the average radius $\langle r \rangle = \int d^3{\bm r}d^3{\bm p}/(2\pi)^3 r W({\bm r},{\bm p})$. We consider all open charmed-hadron states, including $D^\pm, D^0, D_s,\Lambda_c,\Sigma_c,\Xi_c,\Omega_c$ and their excited states with a mass cut of $m<2.6$ GeV for mesons and $m<3.1$ GeV for baryons. Details of the number of states involved in this study and their average radii are provided in the Supplemental Material.

The hadronization of charm quarks via the coalescence process~\eqref{coalescence} is implemented using a Monte Carlo test-particle method, which guarantees the charm quark number conservation naturally. The charm quark momentum at the hadronization hypersurface is determined by the Langevin equation~\eqref{langevin}, and light quarks are assumed to be thermalized with the distribution $f_i(r_i,p_i) = g/(e^{u_\mu p^\mu_i/T} + 1)$, where the local fluid velocity $u_\mu(x_i)$ and temperature $T(x_i)$ are provided by the hydrodynamics. In the calculation we take the quark masses $m_{u,d} = 0.2$ GeV, $m_s = 0.3$ GeV, and $m_c = 1.5$ GeV. These values are slightly smaller than the constituent quark masses due to the chiral symmetry partially restoration around the phase transition boundary, as shown in the Nambu-Jona-Lasinio (NJL) model~\cite{Klevansky:1992qe}.

\begin{figure}[!htb]
	\centering
	\includegraphics[width=0.45\textwidth]{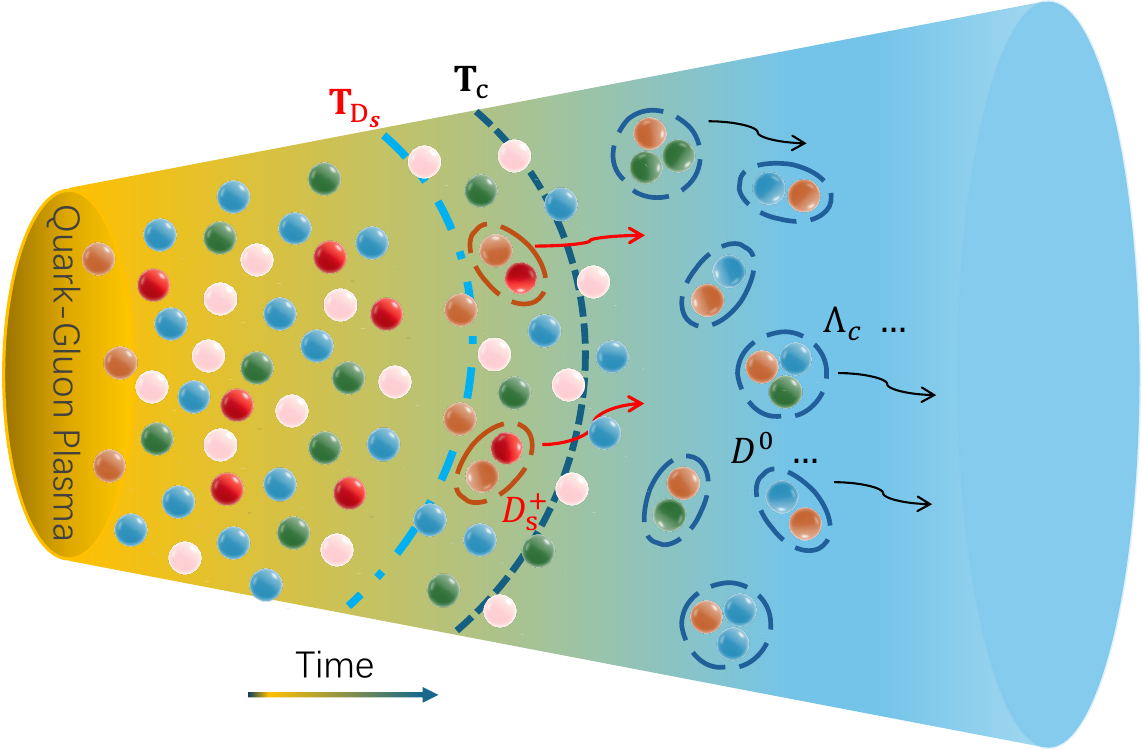}
	\caption{Schematic illustration of sequential hadronization. The dashed and dotted lines represent the hadronization hypersurfaces for the $D_s$ meson at temperature $T_{D_s} > T_c$ and for other hadrons at $T_c$.}
	\label{fig1}
\end{figure}

In the simultaneous coalescence mechanism, all charm quarks hadronize at the same hypersurface characterized by the phase transition temperature $T_{\rm c}$. For the sequential coalescence, charmed hadrons form at different temperatures according to their different binding energies. By solving the 2- and 3-body Dirac equations with in-medium potential, the binding energy $\epsilon(D_s)$ is clearly larger than $\epsilon(D^0)\simeq\epsilon(\Omega_c)\simeq\epsilon(\Xi_c)\simeq\epsilon(\Lambda_c)$~\cite{Shi:2019tji}, corresponding to the hadronization temperature hierarchy 
\begin{equation}
T_{D_s} \simeq 1.2 T_c >T_{D^0}, T_{\Omega_c}, T_{\Xi_c}, T_{\Lambda_c}\simeq T_c.
\end{equation}
Therefore, there are two hadronization hypersurfaces, as schematically shown in Fig.~\ref{fig1}. The first hypersurface is for $D_s$ mesons with formation temperature $T_{D_s} = 1.2 T_c$, and the other is for all the other mesons with formation temperature $T_c$.  

\begin{figure}[!htb]
	\centering
	\includegraphics[width=0.48\textwidth]{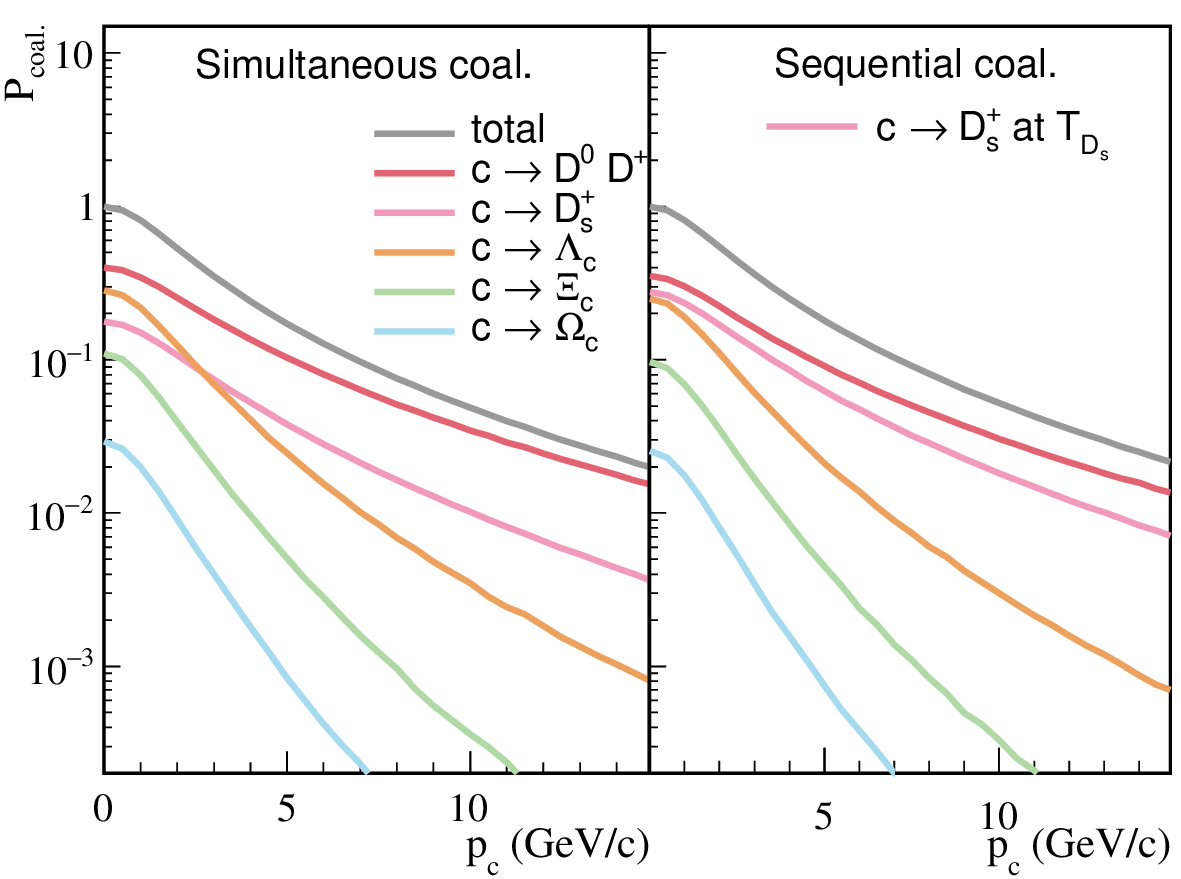}
	\caption {Charm-quark hadronization probabilities for different channels as functions of charm-quark momentum. The left and right panels correspond to the simultaneous and sequential hadronization scenarios, respectively.}
	\label{fig2}
\end{figure}

Fig.~\ref{fig2} shows the charm quark hadronization probabilities for different branches calculated with the simultaneous (left) and sequential (right) hadronization mechanisms. The total probability is the summation over all the branch probabilities; its normalization to unity at charm quark momentum $p = 0$ comes from the fact that at extremely low momentum, the coalescence is the only source of hadronization, and there is no contribution from the fragmentation. The main difference between the left and right panels is the probability for the $D_s$ meson. In sequential hadronization, $D_s$ formation occurs earlier than that of other hadrons. As a consequence of the charm quark number conservation, all the charm quarks are involved in the $D_s$ formation, but fewer charm quarks in the other hadrons formation. This leads to a relative enhancement of the $D_s$ meson in the sequential hadronization, see the difference between the two pink lines in the left and right panels.  

In the coalescence approach, when a charm quark with momentum ${\bm p}$, controlled by the Langevin equations~\eqref{langevin}, reaches the coalescence hypersurface at $T_{D_s}$, its hadronization probability to $D_s$ is $P_{D_s}$. If it does hadronize, then its evolution stops and it transforms into a $D_s$. Otherwise, it continues to propagate in the QGP and hadronizes into other hadrons on the hypersurface at temperature $T_c$, with a second hadronization probability of $1-P_{D_s}$. After the two hadronizations at $T_{D_s}$ and $T_c$, if the charm quark still survives, we let it hadronize through fragmentation, which is governed by the Peterson fragmentation function~\cite{Peterson:1982ak} with the parameter $\epsilon = 0.01$ for charmed mesons and $\epsilon = 0.02$ for charmed baryons~\cite{Das:2016llg}. The fragmentation fractions of a charm quark into various charmed hadrons are extracted from the experimental data~\cite{Lisovyi:2015uqa}. 

\begin{figure}[!htb]
	\centering
	\includegraphics[width=0.48\textwidth]{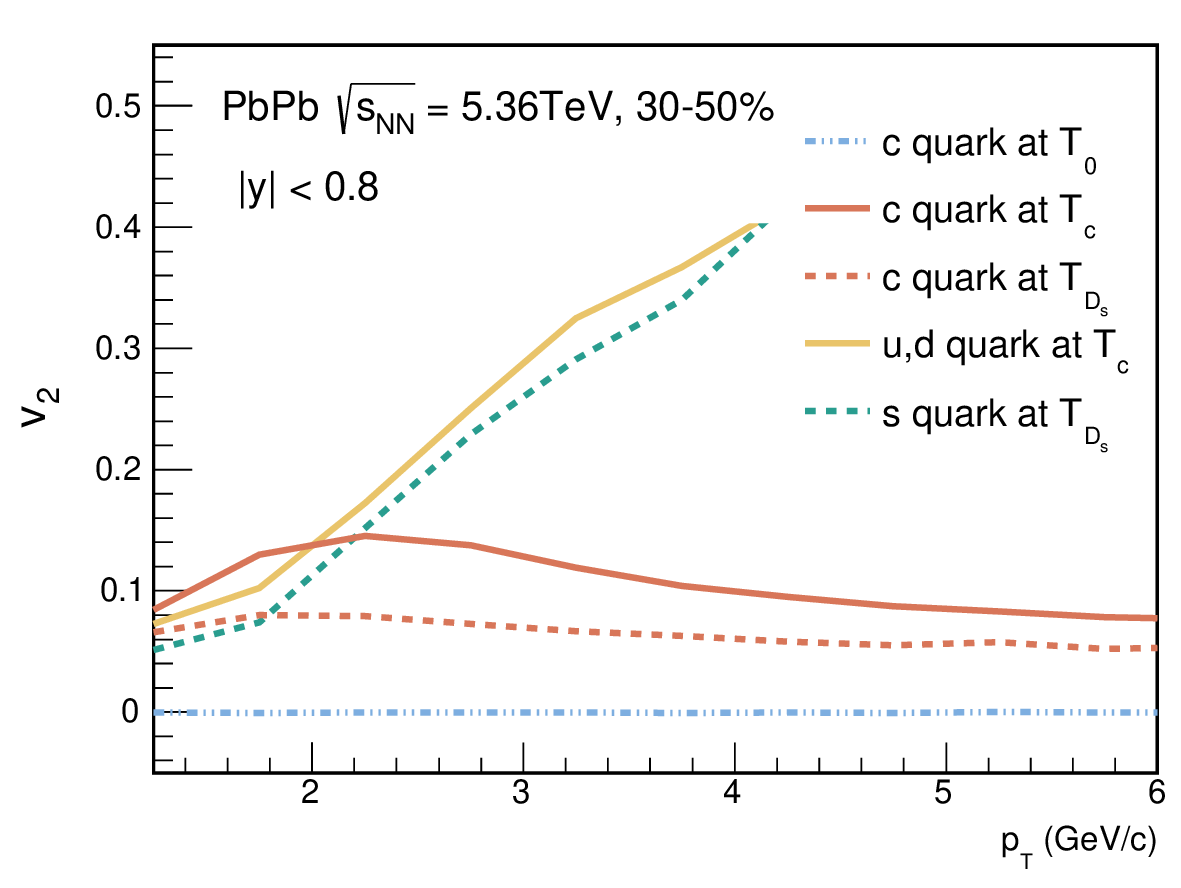}
	\caption{Charm-quark elliptic flow as a function of transverse momentum in $(30$--$50)\%$ central Pb+Pb collisions at center-of-mass energy $\sqrt{s_{\rm NN}}=5.02$ TeV. The blue dot-dashed, red dashed, and red solid lines represent the charm-quark flow at the initial stage and at the two hadronization temperatures $T_{D_s}$ and $T_c$, respectively. For comparison, the light-quark and strange-quark $v_2$ at $T_c$ and $T_{D_s}$ are shown by the orange solid and green dashed lines, respectively.}
	\label{fig3}
\end{figure}

\begin{figure}[!htb]
	\centering
	\includegraphics[width=0.45\textwidth]{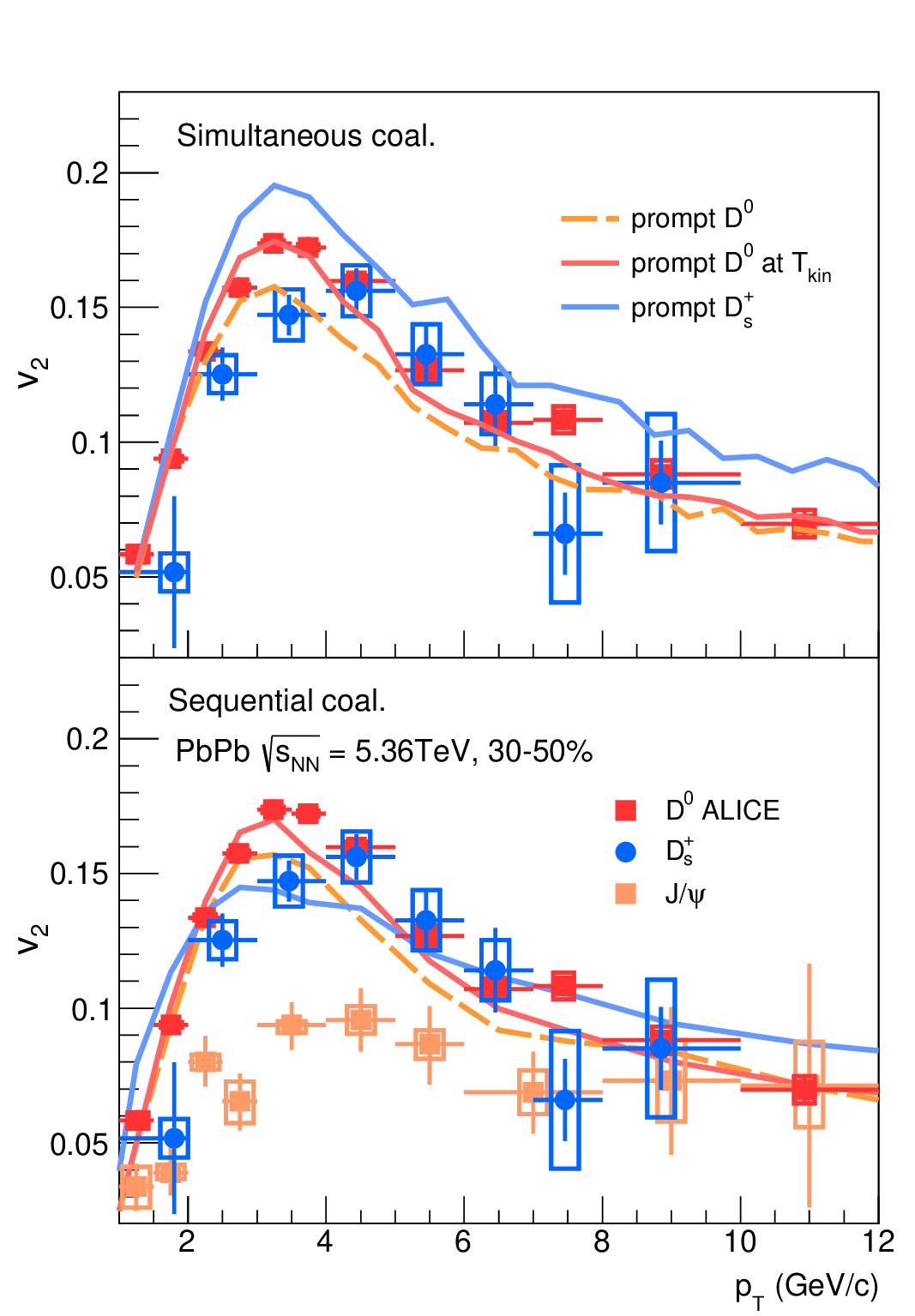}
	\caption{$D$-meson elliptic flow as a function of transverse momentum in $(30$--$50)\%$ central Pb+Pb collisions at center-of-mass energy $\sqrt{s_{\rm NN}}=5.36$ TeV. The theoretical results are obtained in the simultaneous (upper panel) and sequential (lower panel) hadronization scenarios. The red solid and orange dashed lines denote the $D^0$ $v_2$ with and without hadronic-phase rescattering, respectively, and the blue solid line denotes the $D_s$ $v_2$. The experimental data are from the ALICE Collaboration~\cite{Torres:2025cic,ALICE:2026zcz,ALICE:2020pvw}; the $J/\psi$ data are at forward rapidity, $2.5<y<4$.}
	\label{fig4}
\end{figure}

\emph{Flow sequence.--} The particle elliptic flow $v_2$ is initiated from the medium pressure built up in the early stage of the QGP formation, and it grows during the evolution of the medium. Since charm quarks are not the constituents of the QGP, their elliptic flow comes from the interaction with the medium, shown on the right-hand side of the Langevin equations~\eqref{langevin}. The quark flow is then inherited by charmed mesons through the coalescence process. To isolate the in-medium effect at the parton level, we show first in Fig.~\ref{fig3} the charm quark $v_2$ at the initial temperature and the two hadronization temperatures $T_{D_s}$ and $T_c$. The initial $v_2$ is zero, but it grows rapidly due to the strong interaction with the medium and reaches remarkable values at $T_{D_s}$ and $T_c$, especially at low and intermediate transverse momentum. 
Owing to the longer in-medium propagation time of heavy quarks before reaching the $T_c$ hypersurface relative to the $T_{D_s}$ hypersurface, the accumulated charm-quark $v_2$ at $T_c$ is noticeably larger than at $T_{D_s}$. This leads to the $v_2$ splitting for charmed mesons, discussed in the following. For comparison, Fig.~\ref{fig3} also shows the elliptic flow of light quarks $u$ and $d$ at $T_c$ and of strange quarks at $T_{D_s}$; the difference is small.

Recently, the ALICE Collaboration measured the elliptic flow of $D$ mesons; the preliminary data for Pb+Pb collisions in the $(30$--$50)\%$ centrality bin at $\sqrt{s_{\rm NN}}=5.36$ TeV are shown in Fig.~\ref{fig4}. It is clear that in the intermediate transverse momentum region around $p_T=3$ GeV the $D_s\ v_2$ is remarkably smaller than the $D^0\ v_2$. To make our model calculation comparable with the experimental data, we should include the $D$ meson rescattering in the hadron phase. This can be simulated via the Langevin equation. The rescattering strength in the hadron phase is substantially weaker than that in the QGP phase, and the diffusion coefficient $2\pi T{\cal D}_s$ exhibits a significant variation with $T/T_c$. We adopt the temperature dependence of the diffusion coefficient from Ref.~\cite{Torres-Rincon:2021yga} with the kinetic freeze-out temperature $137$ MeV~\cite{Pang:2018zzo}. 
The diffusion of $D_s$ mesons in the hadronic phase is neglected due to their small scattering cross sections. This assumption is supported by experimental evidence indicating that multistrange hadrons decouple near the phase boundary, and is consistent with previous studies~\cite{He:2012df}. The result calculated in the simultaneous hadronization frame and the comparison with the data are shown in the upper panel of Fig.~\ref{fig4}. While the rescattering in the hadron phase increases the $D^0$ flow, the $D^0\ v_2$ is still smaller than the $D_s\ v_2$, which contradicts the experimental data in the intermediate transverse momentum region. 
Detailed analysis indicates that the coalescence component yields a similar elliptic flow $v_2$ for both $D^0$ and $D_s$ mesons under the scenario of simultaneous hadronization. The primary difference stems from the significantly larger fragmentation contribution for the $D^0$ compared to the $D_s$. Since fragmentation typically produces particles with weaker collective flow, it leads to an overall smaller $v_2$ for the $D^0$ meson than for the $D_s$ meson.
The calculation in the sequential hadronization frame is shown in the lower panel of Fig.~\ref{fig4}.
The reversal of the $D_s$ and $D^0$ flow hierarchy in the intermediate $p_T$ region is attributed to the shorter propagation time of charm quarks in the QGP when forming $D_s$ at a higher temperature $T_{D_s}$, compared to $D^0$ hadronization at $T_c$. 
For the high $p_T$ region where the coalescence is not the dominant hadronization mechanism, the $D^0\ v_2$ is still less than the $D_s\ v_2$. We show also the data of $J/\psi\ v_2$ in the lower panel. Since the $J/\psi$ surviving temperature, namely the formation temperature, is much higher than $T_{D_s}$ and $T_c$, its constituents (charm and anti-charm quarks) experience a short propagation time with the QGP, we can then clearly see the $v_2$ hierarchy for the charmed mesons, 
\begin{equation}
v_2(D^0)>v_2(D_s)>v_2(J/\psi).
\end{equation}
We also present the elliptic flow for charmed baryons and the yield spectra and baryon-to-meson ratios; the result is presented in the Supplemental Material~\cite{SM}.

\begin{figure}[!htb]
    \centering
    \includegraphics[width=0.45\textwidth]{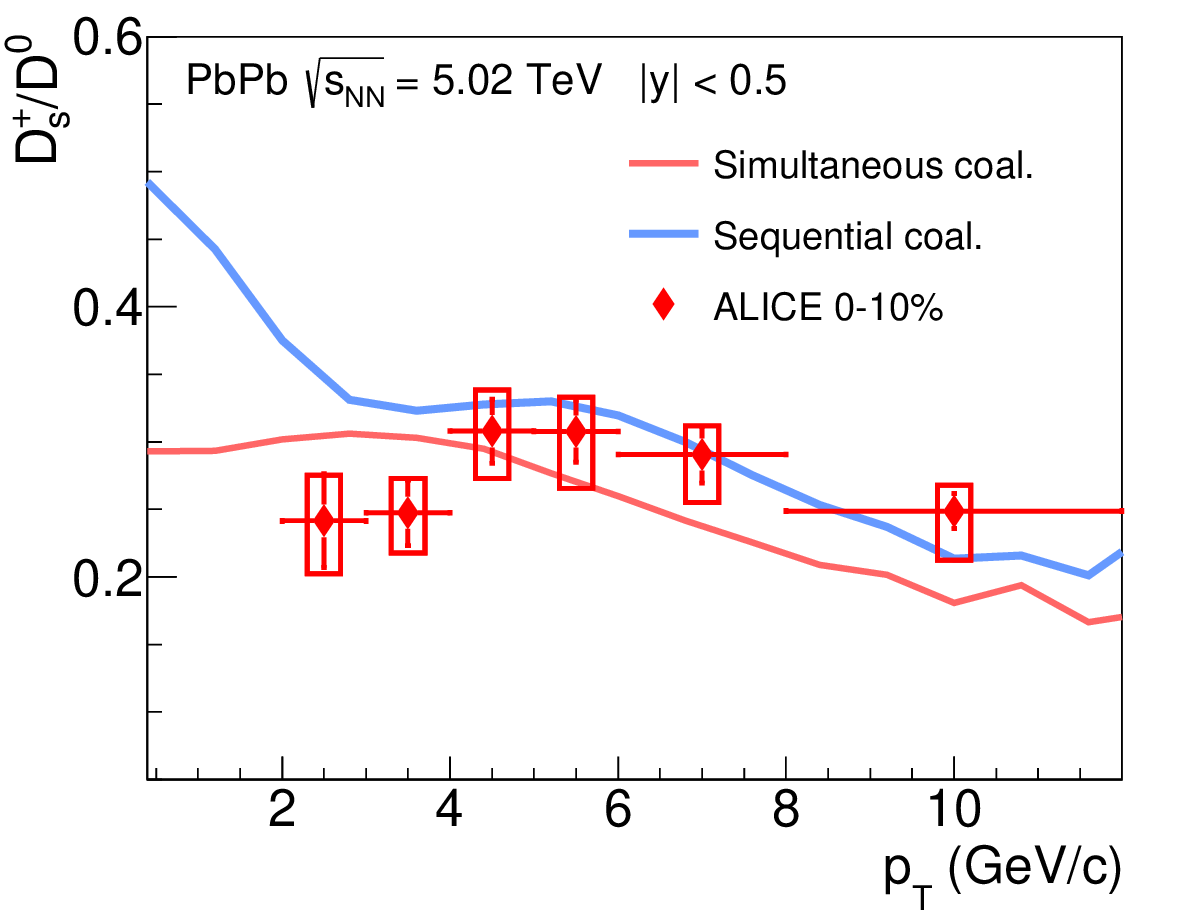}
\caption{The yield ratio $D_s/D^0$ as a function of transverse momentum in $(0$--$10)\%$ central Pb+Pb collisions at center-of-mass energy $\sqrt{s_{\rm NN}}=5.02$ TeV. The theoretical results are obtained in the simultaneous (red line) and sequential (blue line) hadronization scenarios. The experimental data are from the ALICE Collaboration~\cite{ALICE:2021kfc}.}
\label{fig5}
\end{figure}
\emph{Yield ratio.--} We finally examine the yield ratio $D_s/D^0$ as a function of transverse momentum in the simultaneous and sequential hadronization scenarios; the results are shown in Fig.~\ref{fig5}. As shown in Fig.~\ref{fig2}, the coalescence probability for $D_s$ mesons formed earlier is enhanced, and the probability for $D^0$ mesons formed later is suppressed by the charm quark number conservation during the hadronization process. Therefore, it is natural to understand the larger ratio in the sequential scenario (blue line) compared to the simultaneous scenario (red line). In the coalescence dominant region with low and intermediate $p_T$ region, the ratio calculated with the simultaneous scenario tends to be saturated, since $D_s$ and $D^0$ are produced at the same time. However, in the sequential approach, the formation time difference between $D_s$ and $D^0$ becomes more and more important with decreasing $p_T$ and leads to a sharp increase at low $p_T$. While there are no data in the low $p_T$ region, the prediction of a hill, instead of a plateau, in the region of $p_T\lesssim 3$ GeV can be used as complementary evidence of the sequential hadronization mechanisms. In the high $p_T$ region, the ratio shows no significant difference and cannot be reliably distinguished by the current experimental precision.  

\emph{Summary.--} We have studied the sequential coalescence mechanism in relativistic heavy ion collisions and its application to charmed hadron production. We focused on the $D$ meson yields and elliptic flows. In the commonly used simultaneous coalescence scenario, $v_2$ exhibits a positive hierarchy, which means that the $D_s\ v_2$ is always higher than the  $D^0\ v_2$, and the yield ratio $D_s/D^0$ becomes saturated at low $p_T$. In the sequential hadronization approach, the charmed hadron formation temperature $T_h$ or the formation time $t_h$ is controlled by the Dirac equation in the QGP, which leads to a time hierarchy: $t_{J/\psi} < t_{D_s} < t_{D^0}$. In this case, the hadrons produced earlier will be enhanced, and those produced later will be suppressed by the charm quark number conservation, which leads to a hill rather than a plateau of the ratio $D_s/D^0$ at low $p_T$. On the other hand, the longer propagation time of charm quarks in the QGP when forming $D^0$, compared to $D_s$ and $J/\psi$, results in a $v_2$ hierarchy of $v_2(J/\psi) < v_2(D_s) < v_2(D^0)$. These observables constitute direct evidence for sequential hadronization and open a unique window for precisely constraining heavy-quark energy loss and hadronization dynamics.

\textbf{Acknowledgment.--} We thank Xinye Peng for helpful discussions. WD is supported by the National Key Research and Development Program of China with Grant No. 2024YFA1610804, BZ by the National Natural Science Foundation of China with Project No. 12535010, and JZ by the Helmholtz Research Academy Hesse for FAIR. PZ thanks Yantai University under Grant No. 2226001.

\end{document}